\documentstyle[pra,aps,epsfig]{revtex} \twocolumn

\begin{document}

\title{Very long storage times and evaporative cooling of cesium atoms
in a quasi-electrostatic dipole trap}

\author{H.~Engler, T.~Weber, M.~Mudrich, R.~Grimm, and
  M.~Weidem\"uller\thanks{Email: {\sf m.weidemueller@mpi-hd.mpg.de}}}
  
\address{Max-Planck-Institut f\"ur Kernphysik \thanks{Website: {\sf
        http://www.mpi-hd.mpg.de/ato/lasercool}}, 69029 Heidelberg,
    Germany}

\date{\today}

\maketitle

\begin{abstract}
  
  We have trapped cesium atoms over many minutes in the focus of a
  CO$_2$-laser beam employing an extremely simple laser system.
  Collisional properties of the unpolarized atoms in their electronic
  ground state are investigated.  Inelastic binary collisions changing
  the hyperfine state lead to trap loss which is quantitatively
  analyzed.  Elastic collisions result in evaporative cooling of the
  trapped gas from 25\,$\mu$K to 10\,$\mu$K over a time scale of about
  150\,s.


\end{abstract}

The focus of a CO$_2$-laser beam constitutes an almost perfect
realization of a conservative trapping potential for neutral atoms
\cite{takekoshi96:optlett,friebel98:pra}.  Atoms are confined in all
three spatial dimensions by the optical dipole force pointing towards
the maximum of the intensity~\cite{grimm99:adv}.  The CO$_2$-laser
wavelength of $10.6\,\mu$m is far below any optical transitions from
the ground state which has important consequences. As one consequence,
the optical potential becomes quasi-electrostatic, i.e.\ the static
polarizability of the particle determines the depth of the trap ({\em
  quasi-electrostatic trap}, QUEST). Therefore, different atomic
species~\cite{weidemuller99:icols} and even molecules
\cite{takekoshi98:prl} can be confined in the same trap. All substates
of the electronic ground state experience the same trapping potential
in contrast to magnetic traps.  Atoms can thus be trapped in their
absolute energetic ground state which excludes loss through inelastic
binary collisions.  Another consequence of the large laser detuning
from resonance is the negligibly small photon scattering rate so that
heating by the photon momentum recoil does not occur.

In this Rapid Communication we show that storage times of many minutes
can be achieved in a focused-beam dipole trap with a low-cost,
easy-to-use CO$_2$ laser usually employed for cutting and engraving of
materials.  The laser posseses neither frequency stabilization nor
longitudinal mode selection. Despite the simplicity of the laser
system, laser-noise induced heating rates, as first identified by
Savard {\em et al.}~\cite{savard97:pra}, are found to be below
100\,nK/s.  Comparable storage times in a QUEST have recently been
realized by O'Hara {\em et al.}, who utilized an ultrastable,
custom-made CO$_2$ laser~\cite{ohara99:prl}.  The simplicity of our
trap setup in combination with the long storage times are ideal
prerequisites for experiments on interesting collisional properties of
the trapped gas. As one application of the trap, we have studied
hyperfine-changing collisions of unpolarized cesium atoms. As another
important application, evaporative cooling of the trapped gas is
demonstrated, which has so far only once been observed in an optical
dipole trap~\cite{adams95:prl}.

The trapping potential of a CO$_2$-laser beam with a spatial intensity
distribution $I({\bf r})$ is given by $U({\bf r}) = \alpha_{\rm stat}
I({\bf r}) / 2 \varepsilon_0 c$ where $\alpha_{\rm stat}$ denotes the
static polarizability of the atoms
\cite{takekoshi96:optlett,grimm99:adv}. For a focused beam of power
$P$ and waist $w$ one gets a trap potential with a depth $U_0 =
\alpha_{\rm stat} I_0/\varepsilon_0 c$ with $I_0=2 P/\pi w^2$. The
CO$_2$ laser (Synrad 48-2WS) provides 25\,W of power in a nearly
TEM$_{00}$ transversal mode characterized by $M^2 = 1.2$.  The laser
beam is first expanded to a waist of 11\,mm by a telescope, and then
focused into the vacuum chamber by a lens of 254\,mm focal length. The
focus has a waist of 110\,$\mu$m and a Rayleigh range $z_{\rm R}$ of
2.4\,mm, yielding a trap depth 120\,$\mu$K (in units of the Boltzmann
constant $k_{\rm B}$). Gravity lowers the potential height along the
vertical direction to 92\,$\mu$K.  The axial and radial oscillation
frequencies in the harmonic approximation are given by $\omega_{\rm z}
= (2 U_0 / m z_{\rm R}^2)^{1/2}$ and $\omega_{\rm r} = (4 U_0 / m
w^2)^{1/2}$ with $m$ denoting the cesium mass. For our experimental
values one gets an axial oscillation frequency $\omega_{\rm z} / 2\pi
= 8.1$\,Hz and a radial frequency $\omega_{\rm r} / 2\pi = 254$\,Hz.

Atoms are transferred into the dipole trap from a magneto-optical trap
(MOT) containing about $10^6$ atoms. The MOT is loaded from an atomic
beam which is Zeeman-slowed in the fringe fields of the MOT magnetic
quadrupole field. The main vacuum chamber at a background pressure of
about $2 \times 10^{-11}$\,mbar is connected to the oven chamber by a
tube which is divided into differentially-pumped sections to assure a
sufficiently large pressure gradient. To optimize transfer from the
MOT into the dipole trap, the atoms are further cooled and compressed
by decreasing the detuning of the MOT trapping laser from initially $-
2\,\Gamma$ (natural linewidth $\Gamma/2 \pi = 5.3$\,MHz) to $- 20
\Gamma$ with respect to the 6$^2$S$_{1/2}$($F=4$) -
6$^2$P$_{3/2}$($F=5$) transition of the cesium D2 line. After 40\,ms
of compression, the distribution of atoms in the MOT has a rms radius
of 120\,$\mu$m corresponding to an mean density of
$10^{10}$\,atoms/cm$^3$. The temperature is 25\,$\mu$K as measured by
ballistic expansion of the cloud after release from the MOT. After the
laser beams and the magnetic field of the MOT were turned off, the
atoms are trapped in the focus of the CO$_2$ laser beam which was
present during the whole loading phase.  Atoms are prepared in either
the $F=3$ or the $F=4$ cesium hyperfine ground state by shuttering the
MOT trapping laser 1\,ms after or before the MOT repumping laser has
been shuttered, respectively.

\begin{figure}[h]
  \vspace{0cm}
  \centerline{\epsfig{figure=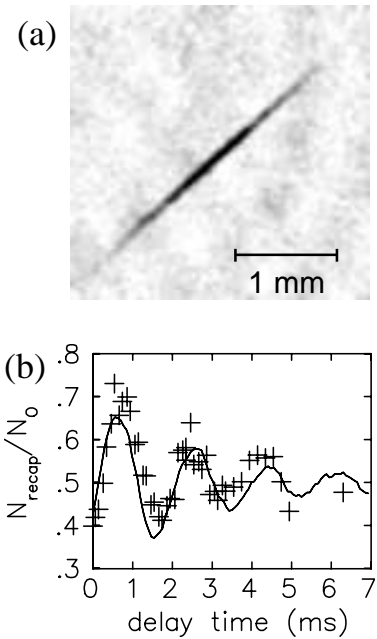,width=5cm}}
\vspace{0.5cm}
  \caption{(a) Distribution of cesium atoms trapped in the focus of a
    CO$_2$ laser beam. The image shows the absorption of a weak
    resonant probe beam. (b) Measurement of the radial oscillation
    frequency of cesium in the dipole trap. Two cycles of
    release-recapture are seperated by a variable delay time.  The
    free expansion time in each cycle is 0.7\,ms. The number of
    recaptured atoms $N_{\rm recap}$ after the second cycle is
    normalized to the number of atoms after the first cycle $N_0$. The
    number of recaptured atoms oscillates at twice the radial
    oscillation frequency.  The result of a Monte-Carlo simulation is
    depicted by the solid line.}
\label{fig:absimage}
\end{figure}

The number and spatial distribution of atoms trapped in the optical
dipole trap are measured by taking an absorption image of the trapped
atoms.  A weak, resonant probe beam of 1\,$\mu$W/cm$^2$ intensity is
pulsed for 100\,$\mu$s and absorption of the atomic cloud is imaged
onto a CCD camera.  Fig.~\ref{fig:absimage}(a) shows a typical absorption
image of atoms trapped in the QUEST. The image is
taken 5\,s after transfer from the MOT. The transmitted intensity
$I_t$ of the probe laser through the atomic sample is described by
$I_t(x,y)=I_0 \exp\left(-A \, \eta(x,y) \right)$ where $I_0$ denotes
the laser intensity and $A$ the absorption cross section for the
resonant transition. The column density $\eta$ is given by the
integral of the density distribution $n(x,y,z)$ along the direction of
the laser beam.

By fitting a thermal equilibrium distribution~\cite{luiten96:pra} to
the data, we derive the mean density $\bar{n}$, the total number of
trapped atoms $N$ and the temperature $T$. The maximum absorption of
the trapped cloud is typically 17\,\% in the center of the
distribution yielding a mean density of
$4\times10^{9}$\,atoms/cm$^3$ in the dipole trap. The atoms have
axially expanded into an rms extension of 750\,$\mu$m while the radial
rms extension is 30\,$\mu$m. The temperature is the same as in the MOT
before transfer indicating that the atoms are cooled into the dipole
trap by the MOT.

Typically $10^5$ atoms are transferred into the dipole trap. Assuming
sufficient ergodicity, the number of atoms transferred from the MOT
into the dipole trap can be determined from the phase-space integral
$\int f_{\rm MOT}({\bf x},{\bf p}) \: \theta (U_0 - \epsilon) \,d{\bf
  x} d{\bf p}$ which describes the projection of the phase-space
distribution $f_{\rm MOT}({\bf x},{\bf p})$ in the MOT onto the
trapping region of the dipole trap. The Heavyside step function
$\theta(U_0-\epsilon)$ equals 1 when the total energy $\epsilon =
U({\bf r})+p^2/2 m$ of atoms in the dipole trap is smaller than the
trap depth, and 0 elsewhere.  Taking our experimental parameters, one
expects $2 \times 10^5$ atoms to be transferred into the dipole trap
in reasonable agreement with the actual value.
  
To measure the radial oscillation frequency of atoms in the dipole
trap, we take advantage of the fast switching capability of the CO$_2$
laser. The laser can be turned off within about 200\,$\mu$s by simply
switching the RF power supply driving the gas discharge. By turning
the laser off for a short time interval (around 1\,ms), the trapped
ensemble moves ballistically until the laser is turned on again
({\em release-recapture}). Part of the atoms will have escaped from the
trap, while the recaptured atoms constitute a non-equilibrium
distribution which oscillates at twice the oscillation frequency. When
the first release-recapture cycle is followed by a second one, the
number of finally recaptured atoms depends on the phase of the
oscillation and thus on the delay time between the two cycles.

In Fig.~\ref{fig:absimage}(b), the number of recaptured atoms is
plotted as a function of delay time between the two release-recapture
cycles.  One observes some cycles of coherent oscillation of the
ensemble until the oscillation dephases mainly due to the
anharmonicity of the potential.  The oscillation period of 1.8\,ms
gives a radial oscillation frequency of 270\,Hz in good agreement with
the value expected for the trap parameters. The solid line in
Fig.~\ref{fig:absimage}(b) shows the result of a Monte-Carlo
simulation of the classical atom trajectories for the sequence of two
release-recapture cycles with variable time delay between them.

The lifetime of atoms in the dipole trap is determined by recapturing
the atoms back into the MOT after a variable storage time. For each
loading and trapping cycle, three camera pictures are taken. The first
picture gives the flourescence of atoms in the MOT shortly before
transfer into the dipole trap. The second one shows the fluorescence
distribution after recapture into the MOT.  The last picture is taken
after the atoms have been released from the recapture-MOT to provide
the background which is then subtracted from the first two images.
The number of atoms is determined from the integral over the
fluorescence image. The particle number determined by this method is
estimated to be accurate within a factor of two. By normalizing the
number of recaptured atoms to the number of atoms initially in the
MOT, fluctuations due to variations in the initial number of atoms can
be cancelled out.  Variations of the particle number in the MOT,
however, are found to be below 10\%, so that normalization was not
used in the data presented here.

\begin{figure}[h]
  \vspace{0cm}
  \centerline{\epsfig{figure=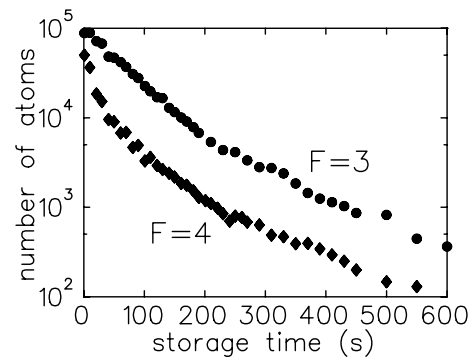,width=6cm}}
\vspace{0.5cm}
  \caption{Evolution of the number of trapped cesium atoms as a
    function of storage time. The atoms are initially prepared in
    either the $F=3$ or $F=4$ hyperfine ground state.}
  \label{fig:lifetime}
\end{figure}

The decay of the number of stored atoms is shown in
Fig.~\ref{fig:lifetime} for the two hyperfine ground states.  Even
after 10\,min storage time, a few hundred atoms can be detected when
prepared in the $F=3$ state.  For atoms in the energetically higher
$F=4$ hyperfine ground state, inelastic binary collisions between
trapped atoms lead to additional trap loss~\cite{weiner1999:rmp}. The
energy of $h \times 9.2$\,GHz released in the collisions is much
larger than the trap depth, therefore both collision partners are
ejected from the trap.

The decay curve for atoms in the $F=4$ state is fitted by the solution
to the differential equation $dN / dt = -\gamma N - \beta \bar{n} N$
where $\gamma$ is the rate for trap loss through collisions with
background gas and $\beta$ is the rate coefficient for inelastic
binary collisions. The fit yields a decay constant $\gamma^{-1} =
165(25)$\,s. This value is consistent with the expected loss rates for
collisions with background gas atoms at a pressure of
$10^{-11}$\,mbar.  No indication for laser-noise induced trap loss is
found. From the fit one finds a decay coefficient $\beta = 2(1) \times
10^{-11}$\,cm$^3$/s with the error being mainly due to the uncertainty
in the absloute particle number. A previous measurement of this
quantity in an opto-electric trap~\cite{lemonde95:epl} gave a similar
result.

Atoms in the energetic ground state $F=3$ can not undergo inelastic
collisions. One would therefore expect a purely exponential decay.
However, the decay curve in Fig.~\ref{fig:lifetime} shows a faster
loss of particles at higher particle numbers. It was checked that more
than 95\% of the particles are initially prepared in the absolute
ground state so that inelastic collisions between trapped particles
can be excluded. At particle numbers below $10^4$, the decay curve
approaches a pure exponential with a decay constant of $\gamma^{-1} =
140(20)$\,s in agreement with the value found for atoms in the $F=4$
state.

The faster initial trap loss can be attributed to evaporation of
high-energetic atoms from the trap leading to
cooling~\cite{ketterle96:adv}. At the initial temperature, the trap
depth $U_0$ is only about $5 k_{\rm B} T$. Therefore there is a
certain probability that atoms leave the trap after an elastic
collision.  The rate for elastic collisions is given by $\bar{n}
\sigma \bar{v}$ with the cross section $\sigma$ and the mean relative
velocity $\bar{v}=4 (k_{\rm B} T/\pi m)^{1/2}$. For evaporative cooling to be
effective, the ratio between the elastic collision rate (providing
thermalization and evaporation) and the rate for inelastic collisions
(causing losses and heating) has to be large.

Up to now, only one experiment is reported on evaporative cooling in
an optical dipole trap~\cite{adams95:prl}.  In order to achieve
sufficiently high densities, Adams {\em et al.}  stored sodium atoms
in a crossed-beam dipole trap which provides tight confinement in all
three dimensions. In the simple focused-beam geometry used in our
experiment it is rather surprising to find evaporative cooling since
the achievable densities are rather low. The reason that evaporation
actually takes place lies in the long storage times of our trap on the
one hand, and the anomalously large elastic cross section of cesium at
low temperatures ~\cite{hopkins00:pra} on the other hand.  Although
the density of cesium atoms in the dipole trap is only of the order of
$10^9$\,atoms/cm$^3$, one expects a thermalization time of only a few
seconds due to the existence of a zero-energy
resonance~\cite{arndt97:prl}. This time scale is indeed more than an
order of magnitude smaller than the storage time.

\begin{figure}[h]
  \vspace{0cm}
  \centerline{\epsfig{figure=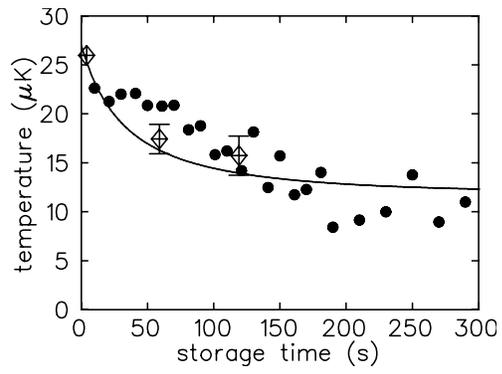,width=6.5cm}}
\vspace{0.5cm}
  \caption{Evaporative cooling of the trapped cesium gas. The dots
    give the temperature derived from the axial extension of the
    trapped atom cloud while the diamonds show the result of a
    ballistic expansion measurement. The solid line depicts a
    simulation of plain evaporative cooling in the trap.}
  \label{fig:evaporation}
\end{figure}

In Fig.~\ref{fig:evaporation}, the temperature of the trapped atoms in
the $F=3$ state is plotted versus storage time. Temperatures derived
from the density distribution shown as the dots are in good agreement
with measurements of ballistic expansion after release from the trap,
which are shown as the three additional points in
Fig.~\ref{fig:evaporation}. One clearly observes cooling of the gas
caused by evaporation of atoms from the trap at constant trap depth
({\em plain evaporation}). The final temperature of about 10$\mu$K
corresponds to roughly 1/10 of the potential depth.  Although the
temperature is reduced by roughly a factor of 2 after 150\,s, the
phase-space density remains almost constant since the particle number
diminishes by a factor of 10 at the same time.  The temperature
evolution for atoms in the $F=4$ state shows a similar behaviour, but
with a slower decrease of the temperature. This is to be expected due
to the faster initial density decrease through inelastic collisions
and the according decrease in the rate of elastic collisions.

We apply a model developed by Luiten {\em et al.}~\cite{luiten96:pra}
to simulate the temperature and particle number evolution during
evaporation. Given the shape of the potential and the corresponding
density of states, the model provides two coupled differential
equations for the evolution of temperature and particle number. For
the true potential function of a focused Gaussian beam, the density of
states diverges as the energy approaches the escape energy of the
trap.  We therefore approximate the trap potential by a
three-dimensional Gaussian $U({\bf r}) \simeq U_0 \exp(-2 (x/w)^2 - 2
(y/w)^2) - (z/z_{\rm R})^2)$ for which the density of states remains
finite.

The potential is fully determined by the trap parameters, therefore
the only adjustable parameter in the model is the the rate $\gamma$
for inelastic collisions with background gas and the cross section
$\sigma$ for elastic collisions.  In the temperature range considered
here, mainly s-wave collisions contribute to the cross section. The
large scattering length of cesium atoms in the $F=3$ ground
state~\cite{hopkins00:pra} results in a temperature-dependent
effective cross section $\sigma(T) = \pi^2 \hbar^2/m k_{\rm B} T$
({\em unitarity limit})~\cite{arndt97:prl}.  Note, that the cross
section in the unitarity limit is completely determined by the
temperature. Since the model explicitly assumes a constant cross
section for elastic collisions~\cite{luiten96:pra}, we have fixed the
effective cross section to the value $\sigma = (930 \,a_0)^2$ ($a_0$ =
Bohr radius) corresponding to a temperature of 15\,$\mu$K.

The result of the simulation for $\gamma^{-1}=130$\,s is shown by the
solid line in Fig.~\ref{fig:evaporation}. The model prediction for the
time scale of temperature decrease is in reasonable agreement with the
experimental data. As can be seen from Fig.~\ref{fig:evaporation},
however, the model slightly overestimates the rate of evaporation.
This discrepancy can be explained by the influence of gravity on the
one hand, and the temperature dependence of the collision cross
section on the other hand. Both effects are not contained in the
model, adn are difficult to be included.  Due to gravity, evaporation
predominantly takes place in only one spatial dimension which slows
down the cooling process~\cite{pinkse98:pra}.  The
temperature-dependent cross section gives rise to an increased
thermalization time~\cite{arndt97:prl} so that the rate of evaporative
cooling is decreased. Nevertheless, the final temperature of the
evaporation process is correctly reproduced by the model. The model
also shows that the initial faster loss of particles in the
$F=3$-state (see curve in Fig.~\ref{fig:lifetime}) mainly stems from
the evaporation process.

Employing a low-cost, off-the-shelf CO$_2$ laser without any
longitudinal mode selection, our major concern was possible heating by
laser-noise induced fluctuations of the trap potential.  From the
observation of evaporative cooling with a temperature decrease of
about $10\,\mu$K over a time scale of 100\,s, one can infer that
heating rates by laser noise are much lower than 100\,nK/s. From the
storage times of several minutes one can infer a similar upper bound
for the laser-noise induced heating rates.

Due to its simple ingredients, our setup represents a minimalistic
version of an optical dipole trap providing very long storage times.
The universality of the QUEST opens way to experiments aiming at
fundamental questions. Atomic spin states can be prepared in the QUEST
with very long coherence times which is of importance for applications
in quantum computing and for the search of a permanent electric dipole
moment of atoms. Another intriguing prospect is the formation and the
storage of cold molecules.  In the same dipole trap, we have recently
also trapped lithium atoms achieving similar storage times as for
cesium~\cite{weidemuller99:icols}.  Currently, we are studying the
properties of a simultaneously stored mixture of both species in view
of sympathetic cooling and photoassociation of cold heteronuclear
dimers.

We gratefully acknowledge contributions of M.~Nill in the early stage
of the experiment and stimulating discussions with A.~Mosk. We are
indebted to D.~Schwalm for encouragement and support.

\end{document}